\documentclass[twocolumn,showpacs,superscriptaddress,pra]{revtex4-1}
\usepackage{graphicx}
\usepackage{times}
\usepackage{textcomp}
\usepackage{amsmath,amssymb}
\usepackage{color}
\begin{document}
\title{Quenching and generation of random states in a kicked Ising model}
\author{Sunil K. Mishra}
\email{iitk.sunilkm@gmail.com}
\affiliation{Institut f\"{u}r Physik, Martin-Luther-Universit\"{a}t Halle-Wittenberg, 06120 Halle, Germany}
\author{Arul Lakshminarayan}
\email{arul@physics.iitm.ac.in}
\affiliation{Department of Physics, Indian Institute of Technology Madras, Chennai -600036, India}
\begin{abstract}
The kicked Ising model with both a pulsed transverse and a continuous longitudinal field is studied numerically. 
Starting from a large transverse field and a state that is nearly an eigenstate, the pulsed transverse field is 
quenched with a simultaneous enhancement of the longitudinal field. The generation of multipartite entanglement is observed 
along with a phenomenon akin to quantum resonance when the entanglement does not evolve for
certain values of the pulse duration. Away from the resonance, the longitudinal field can drive the entanglement 
to near maximum values that is shown to agree well with those of random states. Further evidence is presented that 
the time evolved states obtained do have some statistical properties of such random states. For contrast the case when the 
fields have a steady value is also discussed. 
 \end{abstract} 
\newcommand{\newc}{\newcommand}
\newc{\beq}{\begin{equation}}
\newc{\eeq}{\end{equation}}
\newc{\kt}{\rangle}
\newc{\br}{\langle}
\newc{\beqa}{\begin{eqnarray}}
\newc{\eeqa}{\end{eqnarray}}
\newc{\pr}{\prime}
\newc{\longra}{\longrightarrow}
\newc{\ot}{\otimes}
\newc{\rarrow}{\rightarrow}
\newc{\h}{\hat}
\newc{\bom}{\boldmath}
\newc{\btd}{\bigtriangledown}
\newc{\al}{\alpha}
\newc{\be}{\beta}
\newc{\ld}{\lambda}
\newc{\sg}{\sigma}
\newc{\p}{\psi}
\newc{\eps}{\epsilon}
\newc{\om}{\omega}
\newc{\mb}{\mbox}
\newc{\tm}{\times}
\newc{\hu}{\hat{u}}
\newc{\hv}{\hat{v}}
\newc{\hk}{\hat{K}}
\newc{\ra}{\rightarrow}
\newc{\non}{\nonumber}
\newc{\ul}{\underline}
\newc{\hs}{\hspace}
\newc{\longla}{\longleftarrow}
\newc{\ts}{\textstyle}
\newc{\f}{\frac}
\newc{\df}{\dfrac}
\newc{\ovl}{\overline}
\newc{\bc}{\begin{center}}
\newc{\ec}{\end{center}}
\newc{\dg}{\dagger}
\newc{\prh}{\mbox{PR}_H}
\newc{\prq}{\mbox{PR}_q}
\newc{\tr}{\mbox{tr}}
\newc{\pd}{\partial}
\newc{\qv}{\vec{q}}
\newc{\pv}{\vec{p}}
\newc{\dqv}{\delta\vec{q}}
\newc{\dpv}{\delta\vec{p}}
\newc{\mbq}{\mathbf{q}}
\newc{\mbqp}{\mathbf{q'}}
\newc{\mbpp}{\mathbf{p'}}
\newc{\mbp}{\mathbf{p}}
\newc{\mbn}{\mathbf{\nabla}}
\newc{\dmbq}{\delta \mbq}
\newc{\dmbp}{\delta \mbp}
\newc{\T}{\mathsf{T}}
\newc{\J}{\mathsf{J}}
\newc{\sfL}{\mathsf{L}}
\newc{\C}{\mathsf{C}}
\newc{\B}{\mathsf{M}}
\newc{\V}{\mathsf{V}}
\maketitle
Quantum entanglement has been a widely studied feature of quantum mechanics \cite{nielsen}. 
Of late this study has been substantially enhanced as it finds a central role in various quantum 
information protocols such as teleportation \cite{nielsen,bennett,peres}. In the last few years entanglement has also been 
studied in many condensed matter contexts, especially for various  types of spin-chains \cite{bose,subram1,latorre,osborne,amico,diptiman}. 
However, much of the work has focused around integrable models such as the transverse field Ising model, 
the Heisenberg model etc., with relatively little light being shed on the non-integrable ones \cite{arul1,arul2,casati}.
Quantum Ising spins under the application of a magnetic-field in the transverse direction present various 
interesting features while crossing across the quantum critical point \cite{sachdev}. The spin dynamics has been studied
by driving the system out of its equilibrium by quantum quenching. In a typical quench dynamics study the Hamiltonian of the 
system $\mathcal{H}(\lambda)$ at $t=0$ depends on the parameter $\lambda$, that is itself is a function
of time. The evolution of the state of the system in
this case becomes very interesting as the system is out of the stationary state behavior. 
Recent progress in spin chain studies point out the role of entanglement as a method to characterize the quantum phase 
transitions \cite{osborne,amico,diptiman,tribedi}. 
These studies show the qualitative change of various entanglement measures near or at the Quantum critical point, in turn 
identifying and characterizing quantum phase transitions. Keeping these studies in mind one can
also speculate qualitative changes in various entanglement measures as one crosses across the 
quantum critical point using quench dynamics in the nonintegrable regime.

In the present manuscript we will discuss some results for Ising spins under the influence of time varying
and {\it tilted} magnetic fields. In the section \ref{model1} we will start the manuscript by defining some important 
entanglement measures which will be used to quantify entanglement. In the subsequent section Section \ref{results}, 
 we discuss the results of the model extensively. The manuscript is summarized in section \ref{conclude}.     
\section{Model}
\label{model1}
In the present study we start with a kicked Ising model which is a variant of the transverse field Ising model. The usual Ising model Hamiltonian with 
both a transverse and a longitudinal magnetic field is given by:
\begin{eqnarray}
 \mathcal{H}_{TL}=J_z\sum_{j=1}^L \sigma_j^z \sigma_{j+1}^z +h_z\sum_{j=1}^L \sigma_j^z+h_x\sum_{j=1}^L \sigma_j^x
\label{isin_hal}
\end{eqnarray}
where $J_z$ is the exchange interaction strength, $h_x$ is the external transverse field and $h_z$ is the external longitudinal field. 
The presence, simultaneously, of longitudinal and transverse field terms leaves the model non-integrable.
 The above model in the absence of the longitudinal field has been widely studied using Jordan-Wigner transformation \cite{jordan,sachdev}
where the spin system is mapped to a system of noninteracting fermions. However, inclusion of the longitudinal field 
term terminates the liberty to map the problem into noninteracting fermions due to the presence of 
non-quadratic fermionic operators after the transformation. 
Hence this study will rely mostly on numerical studies and calculate various entanglement measures using such calculations.
 The inclusion of longitudinal field gives rise to many peculiar phenomena in the quench dynamics, and this concentrates on a few of them.  
The variant of the Ising model which will be studied in the present manuscript involves kicks of the transverse magnetic
field at regular intervals.

 The Hamiltonian in this case is
\begin{eqnarray}
 \mathcal{H}(t)&=&J_z\sum_{j=1}^L \sigma_j^z \sigma_{j+1}^z + h_z(t)\sum_{j=1}^L \sigma_j^z \nonumber \\
&+&
 \sum_{k=-\infty}^{\infty}\delta(k-\frac{t}{\tau}) h_x(t)\sum_{j=1}^L \sigma_j^x.
\label{mod}
\end{eqnarray}
For the above kicked Hamiltonian, the time evolution between two successive kicks is that of a simple Ising model with longitudinal field. At the
kick, the transverse field spike is strong which makes the interaction term unimportant. In the absence of the longitudinal field
the model becomes integrable. 
  In all the studies below, we set $J_z=1$ and most of the time consider periodic boundary conditions i.e., $\sigma_{L+1}=\sigma_1$. The setting of $J_z=1$ fixes
 one time-scale $1/J_z$ as unity and all other times to be chosen are in these units.
    
The transverse field is swept from a maximum value $h_{x0}$ with a rate of quenching $\alpha$. $\tau$ is 
the time between two kicks. The introduction of this additional time scale leads to interesting consequences, but it is also convenient as
a numerical scheme as when $\tau \ll 1$ the Hamiltonian $\mathcal{H}$ tends to $\mathcal{H}_{TL}$ in Eq.~(\ref{isin_hal}), possibly with time varying fields.

The operator that evolves states of the system from one application of the transverse field to the next 
is the quantum map \cite{balazsberryvorostabor}. 
The quantum map, which is the propagator evaluated between $k\tau^+$ and $(k+1)\tau^+$ (where $t^+$ indicates a time just 
after $t$) is given by $U(k)=U_x(k) \, U_z(k)$ where
\begin{equation}
 U_z(k)=\prod_{j=1}^{L_0} \exp\biggl(-i\tau \sigma_j^z\sigma_{j+1}^z-\int_{k\tau}^{(k+1)\tau} h_z(t) dt \sigma_j^z\biggr)
 \label{Eq:Uz}
\end{equation}
and
\begin{equation}
 U_x(k)=\prod_{j=1}^{L_0} \exp(-i\tau h_x(k\tau) \sigma_j^x).
\end{equation}
Here $L_0= L$ or $L-1$  for periodic or open chains respectively.
Note that due to the generally aperiodic time dependence, the quantum map so constructed is itself non-autonomous, and the dependence on $k$, the 
kick number, is a result of this.
As the terms containing $\sigma_x$ and $\sigma_z$ do not commute, the quantum map $U(k)$ is different from $\exp\bigl(-i\int_{0}^{\tau} \mathcal{H}_{TL}(t)dt\bigr)$, where $\mathcal{H}_{TL}$ is the autonomous Hamiltonian in Eq. ~(\ref{isin_hal}), even in the case when $h_{x,z}(t)$ are independent of time. 
It has been shown that the kicked transverse Ising model with $h_z=0$ is integrable \cite{prosen}, and has been solved in \cite{arul1} using Jordan-Wigner transformation.

The state of the system at any time $t=N\tau^+$ is given by
\begin{eqnarray}
 \vert\psi(t)\rangle=\prod_{k=0}^{N-1} U(k)\vert\psi(0)\rangle,
\end{eqnarray}
where $\vert\psi(0)\rangle$ is the initial state at time $0^+$, and the product of non-commuting operators $\{U(k)\}$ is time-ordered from right to left. 
The quantum map $U(k)$ can be considered as a series of quantum gates $U_z(k)$ and $U_x(k)$ on the pair of nearest-neighbor qubits
and on the individual qubits.
 In this work we will mostly discuss the nonintegrable
model where a time dependent longitudinal field acts on the system. For the most part we use sinusoidally varying 
magnetic fields $h_x(t)=h_{x0}\cos \alpha t$ and $h_z(t)=h_{z0}\sin \alpha t$. 

The following entanglement measures are studied: the concurrence $C(i,j)$ which is the
entanglement of spin $i$ with spin $j$. This is based on the two-spin reduced density matrix $\rho_{ij}$. The $Q$ measure
is given by 
\begin{eqnarray}
 Q=2\biggl(1-\frac{1}{L}\sum_{k=1}^L{\rm Tr}(\rho_k^2)\biggr),
\end{eqnarray}
 where $\rho_k$ is the one-spin reduced density matrix of the $k^{\rm{th}}$ spin obtained by tracing out the rest.
The block-entropy $S_{L/2}$ is given by 
\begin{eqnarray}
 S_{L/2}=-{\rm Tr}_{1,\cdots,L/2}[\rho_{1,\cdots, L/2}\log_2(\rho_{1,\cdots,L/2})],
\end{eqnarray}
  where,
\begin{eqnarray}
 \rho_{1,\cdots,L/2}={\rm Tr}_{L/2+1,\cdots, L}(|\psi\rangle\langle\psi|).
\end{eqnarray}
It is believed that $Q$ and the entropy $S_{L/2}$ are multipartite measures of entanglement and will therefore
be complementary to the concurrence, in the sense that we can expect states that maximize concurrence to have low
$Q$ and entropy. Generic (random) states have a vanishing probability of a nonzero $C(i,j)$ and a nearly maximal entropy and $Q$ values \cite{scottcaves}.
The eigenvalues of the reduced density matrix $\rho_{1,...,L/2}$ contains interesting information, and we will also study the distribution of these further below.

%
 
\begin{figure}[t]
\includegraphics [angle=0,width=1.0\linewidth] {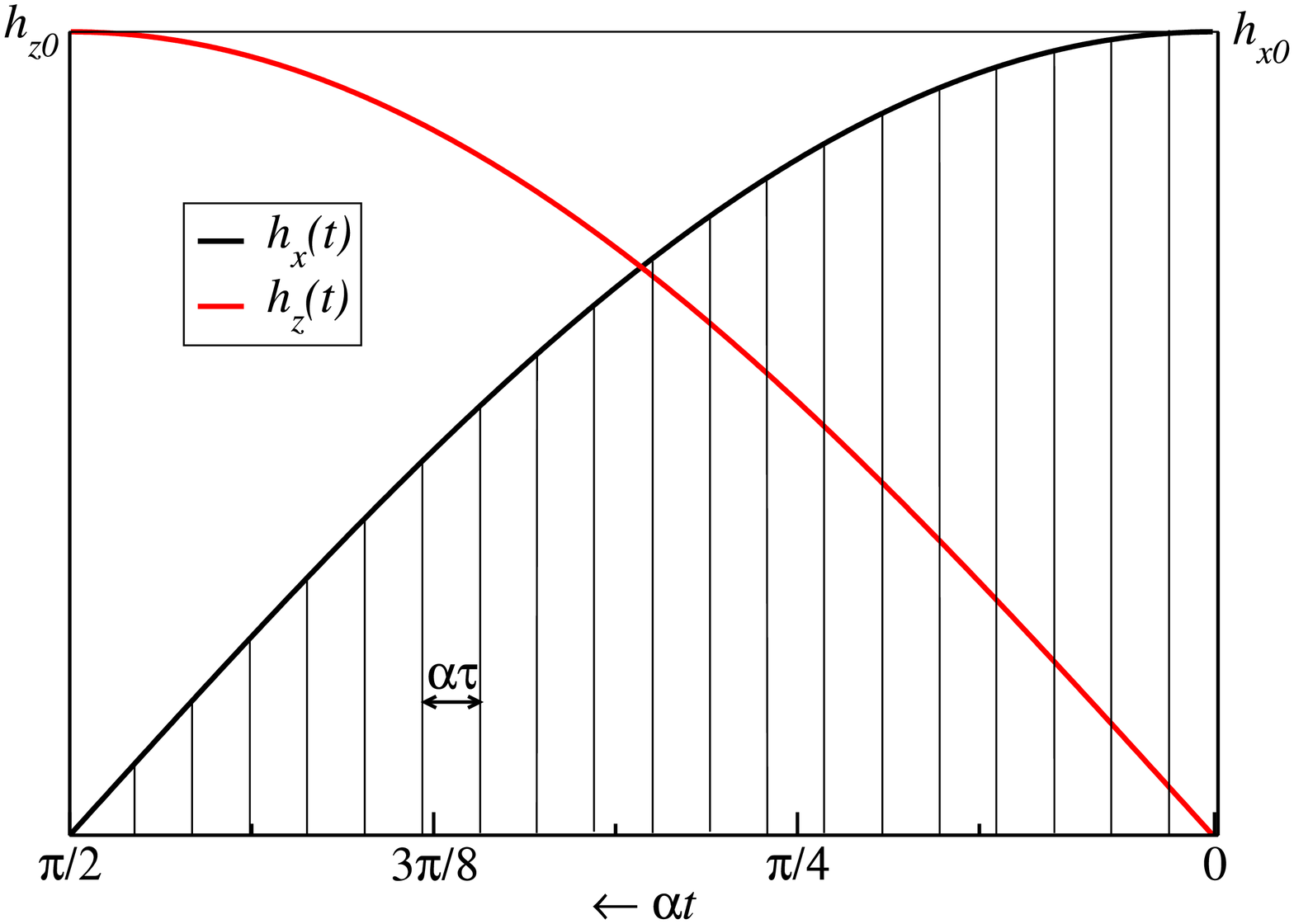}
\caption{The time dependence of transverse and longitudinal fields is shown in the figure. At $t=0$ transverse field is maximum and
longitudinal field is set to zero. The kicks of the transverse field are shown as lines separated by the period $\tau$.}
\label {fields}
\end{figure}
 \begin{figure}[!t]
  \includegraphics[width=\columnwidth]{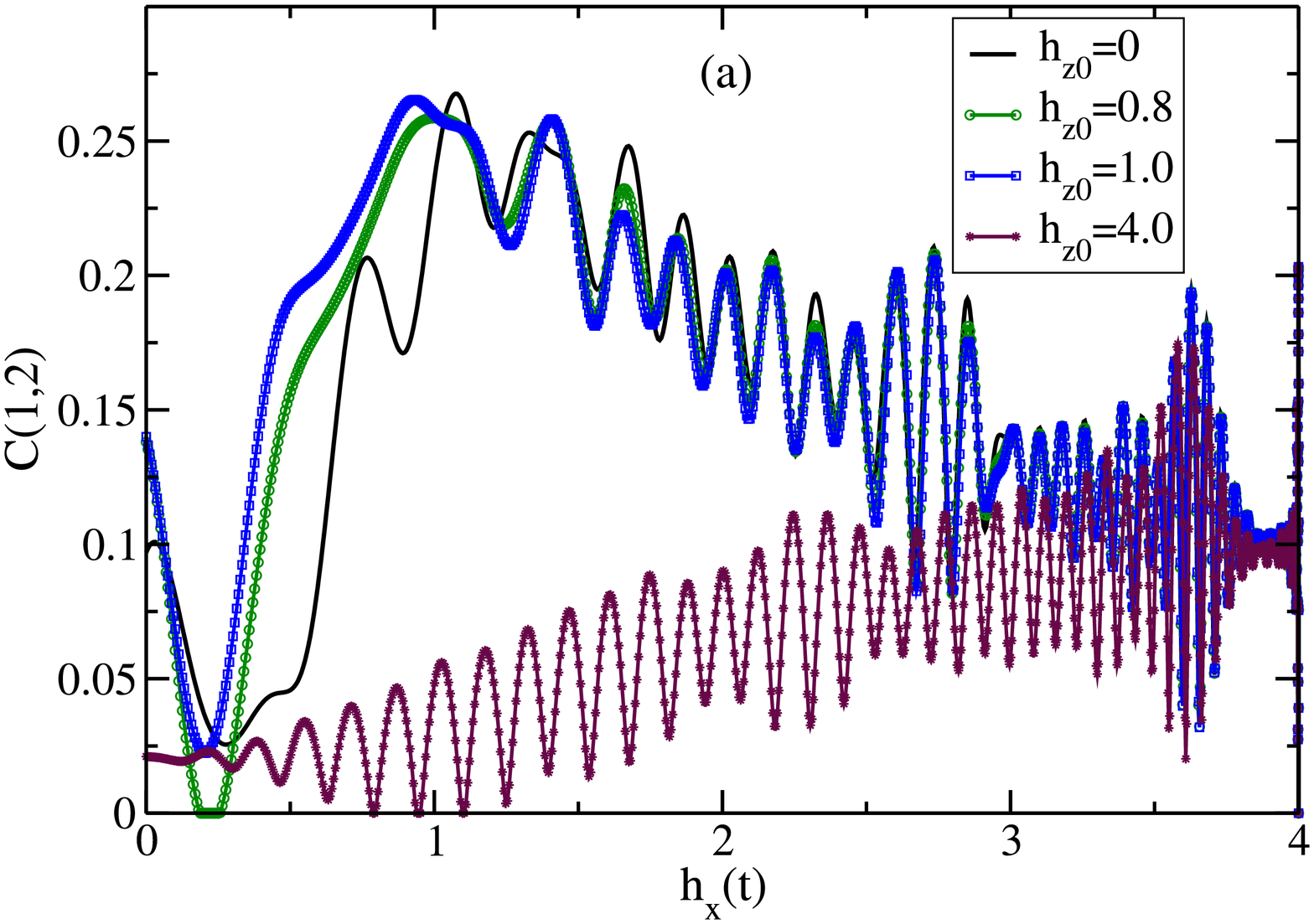} \\
    \includegraphics[width=\columnwidth]{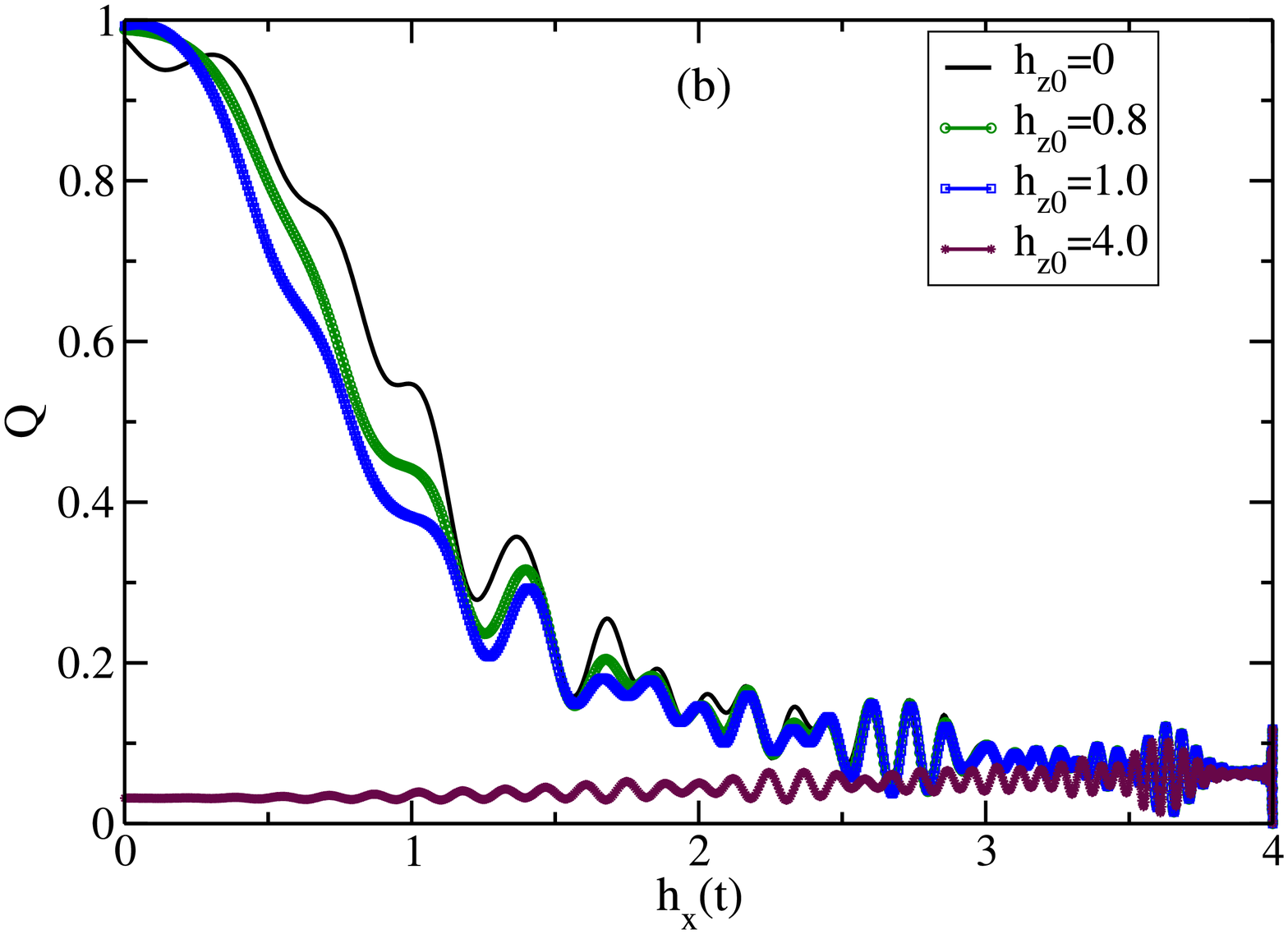} \\
    \includegraphics[width=\columnwidth]{fig2c.eps} 
  \caption{ (a) Nearest-neighbor concurrence, (b) $Q$-measure, and (c) von Neumann entropy is shown with the quenching in the 
  transverse field for various
  values of peak longitudinal fields $h_{z0}$. In all the cases $L=20$, $t_{\rm{max}}=20$ and $\tau=0.02$. The quenching fields are given by 
 $h_z(t)=h_{z0}\sin(\alpha t)$ and $h_x(t)=h_{x0}\cos(\alpha t)$, with $h_{x0}=4.0$ and $\alpha=\pi/2 t_{\rm max}$ where $t_{\rm max}=20$.
 Periodic boundary conditions are used.}
      \label{withhx}
  \end{figure}
\section{Results and discussions }
\label{results}
The model as discussed in Eq.~\ref{mod} is analyzed using sinusoidal longitudinal and transverse magnetic fields of
 frequency $\alpha$: $h_z(t)=h_{z0}\sin(\alpha t)$ and $h_x(t)=h_{x0}\cos(\alpha t)$, see Fig.~(\ref{fields}).
The entanglement measures discussed in the above section are calculated for the state $|\psi(t)\kt $ 
where at $t=0$ all the spins in the chain are in the initial state $\lvert\rightarrow \rangle$ state, 
where $\lvert\rightarrow \rangle$ is the eigenstates of $\sigma_x$ with eigenvalue $+1$. 
\begin{figure}[!t]
\includegraphics [angle=0,width=1.0\columnwidth]{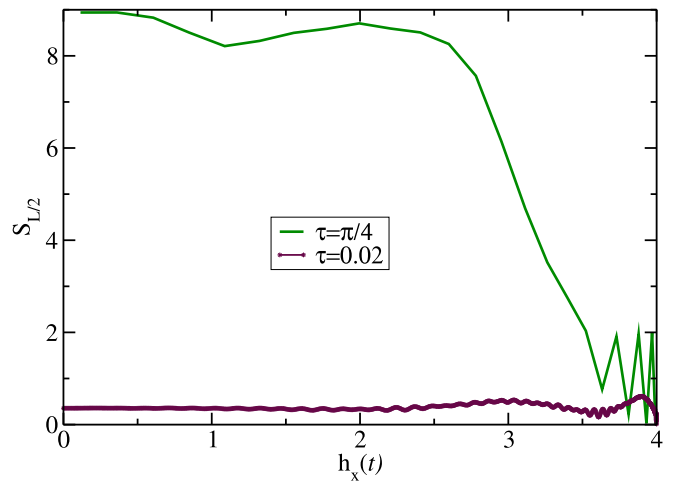}
\caption{   The von Neumann entropy as a function of the quenching transverse field for two very different kick periods.
The quenching fields are given by  $h_z(t)=h_{z0}\sin(\alpha t)$ and $h_x(t)=h_{x0}\cos(\alpha t)$, with $h_{z0}=4.0$, $h_{x0}=4.0$, and 
$\alpha=\pi/2 t_{\rm max}$ where $t_{\rm max}=20$. Periodic boundary conditions are used.}
\label {entropytau}
\end{figure}
Thus just after kick $N$ at time $t=N\tau^+$ the state is
\begin{eqnarray}
 \vert\psi(t)\rangle= \prod_{k=0}^{N-1} U(k) \otimes^L\lvert\rightarrow \rangle
\label{state}
\end{eqnarray}

As time increases the envelope of kicked transverse field $h_x$ decreases from its
 maximum value $h_{x0}$ and during the same time the longitudinal field $h_z$ increases 
 from $0$ by virtue of the cosine and sine functions. 
The frequency of the fields is kept fixed to the value $\alpha=\pi/2t_{\rm max}$, where $t_{\rm max}$ is the 
time spent during the evolution. 
The value of the kick interval $\tau$ is fixed to be $0.02$ and the time evolution for $L=20$ spins is performed till $t_{\rm max}=20$.
Fig.~\ref{withhx} shows the nearest-neighbor concurrence, Q-measure and von Neumann entropy for the quench case
of the kicked model.
In Fig.~ \ref{withhx}(a) The nearest neighbor concurrence versus the quenched transverse field for various 
longitudinal fields is shown. In all the cases we see oscillations in the concurrence value as the transverse field quenches to zero.
As the concurrence is a measure of two-qubit entanglement, we see that for higher value of
peak longitudinal values, the pairwise entanglement vanishes at earlier time (or high transverse field). 
 For the similar cases in Fig.~\ref{withhx}(b), the $Q$ measure of entanglement reaches the maximum value of unity where the concurrence 
vanishes. As the increase in the multipartite entanglement is associated with the
decrease in the average two body entanglement as measured by the concurrence in Fig.~\ref{withhx}(a) we 
see the signature of increased multipartite entanglement with the longitudinal field, and with time. 
In Fig.~\ref{withhx}(c) the von Neumann entropy is shown for different cases as discussed above. The von Neumann entropy 
supports the multipartite nature of the generated entanglement, and is qualitatively similar to the simpler quantity $Q$.

\begin{figure}[!t]
\includegraphics [width=\linewidth] {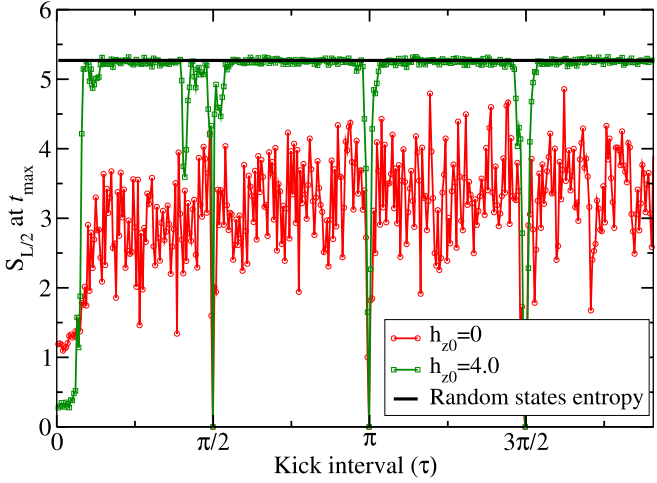}
\caption{The von Neumann entropy at $t_{\rm{max}}$, where the quench is over is plotted against kick interval $\tau$. 
During the quench $h_x(t)$ (and $h_z(t)$) starts from $h_x(t)=4.0$ (and $h_z(t)=0$) and finishes to $h_x(t)=0$ (and $h_z(t)=h_{z0}$).
A comparison between two cases 
with $h_{z0}=0$ and $4$ is shown. In all the cases $t_{\rm{max}}=400$ and $L=12$ with periodic boundary conditions. 
A reference line corresponding to random state entropy is also shown.}  
\label {entropy_rate}
\end{figure}
As the kick interval $\tau=0.02 \ll 1$ is small the $\delta$ function is almost acting continuously and the evolution 
 is to a good approximation that of the Hamiltonian in Eq.~\ref{isin_hal}, with time dependent fields.
The dependence on the time between the kicks will indeed be crucial and this is illustrated in Fig.~(\ref{entropytau}) where the very large 
von Neumann entropy(entanglement) when $\tau=\pi/4$ is to be contrasted to the case when $\tau=0.02$, 
although clearly $\tau \ll t_{\rm max}$ in both cases as $t_{\rm max}=20$, the number of kicks for
$\tau=0.02$ ($N=1000$) are much larger than that ($N=25$) for $\tau=\pi/4$ case. 
 Also the case of $\tau=\pi/4$ has a time scale that of the order of the time-scale set by the 
interaction $J_z$.

In the following we study the Von Neumann entropy at the end of the quench when $h_x(t)=0$, the start corresponding to the
 maximum transverse field $h_x(t)=h_{x0}$. During the process the longitudinal field 
builds up from zero to reach its maximum value. Shown are cases for vanishing ($h_{z0}=0$) and $h_{z0}=4$ peak value of the 
longitudinal field. For every value of $\tau$ taken, there are an integer number $N$ of kicks, such that $N\tau=t_{\rm max}$ is fixed. 
The rate of quench is $\alpha=\pi/2N\tau$. As can be seen in 
Fig. \ref{entropy_rate}, the case when longitudinal field is switched on shows peculiar behavior. 
For an initial state all $\sigma_x$ eigenstates, it can be seen that the entropy increases considerably for 
moderate kicking periods and interestingly sharply vanishes exactly at $\tau =\pi/2,\pi,3\pi/2,...$, and recovers almost immediately 
to very large values.

The reason for this is not hard to find. Define the interaction part in $U_z(k)$ 
of Eq.~(\ref{Eq:Uz}) as $U_z^I (\tau)= \prod_{j=1}^{L_0} \exp \left(-i\tau \sigma_j^z\sigma_{j+1}^z\right)$. Then with periodic boundary 
conditions we have that 
\begin{eqnarray}
U_z^I(\pi/2)=\prod_{j=1}^L  \exp\left(-i\frac{\pi}{2}\sigma_j^z\sigma_{j+1}^z\right)= (-i )^L \mathbb{I}_{2^L},
\label{Eq:UzIdent}
\end{eqnarray}
where $\mathbb{I}_{2^L}$ is the $2^L-$ dimensional identity. This is proved by considering the action of $U_z^I(\pi/2)$ on 
states $|a_1\cdots a_L\kt$ where $a_j \in \{0,1\}$, and $\sigma_j^z |a_j \kt = (1-2a_j)|a_j \kt$, that is these are the eigenstates of
$\sigma^z$. Therefore
\beq
\begin{split}
U_z^I(\pi/2) |a_1\cdots a_L\kt & = e^{-i \f{\pi}{2}\sum_{j=1}^{L}(1-2a_j)(1-2a_{j+1})}|a_1\cdots a_L\kt\\
& e^{-i \f{\pi}{2}L}e^{-i \pi (a_1+a_{L+1})}|a_1\cdots a_L\kt.
\end{split}
\eeq
With periodic boundary conditions, $a_{L+1}=a_{L}$ and hence the phase becomes independent of 
the $\{a_{j}\}$, and as 
the set $\{ |a_1\cdots a_L\kt, \,  a_j \in \{0,1\}\}$ is complete 
the identity in Eq.~(\ref{Eq:UzIdent}) follows.

\begin{figure}[!t]
\includegraphics [angle=0,width=1.0\columnwidth] {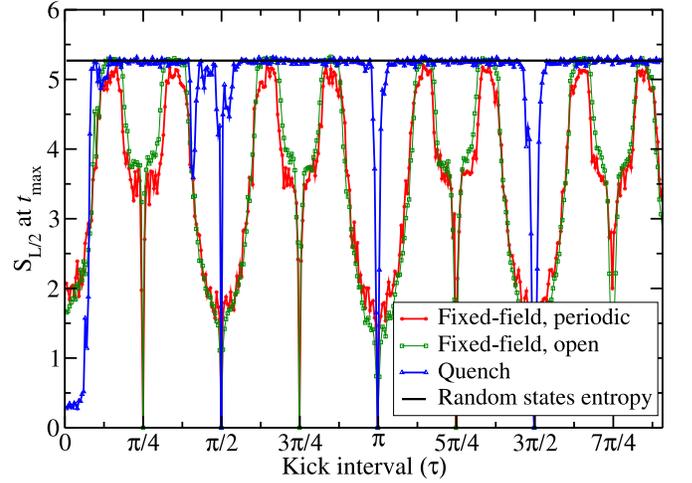}
\caption{ The von Neumann entropy at $t_{\rm{max}}=400$ is plotted against kick interval $\tau$ for various protocols, namely: 	(i) Fixed-field, periodic;
where constant $h_x=2.0$ and $h_z=2.0$ is applied during the process and periodic boundary conditions are taken into account. (ii) Fixed-field, open; where
fixed fields $h_x=2.0$ and $h_z=2.0$ is applied all time with open boundary conditions.(iii) Quench; where
 $h_x(t)$ (and $h_z(t)$) starts from $h_x(t)=4.0$ (and $h_z(t)=0$) and finishes at $t_{\rm{max}}$ to $h_x(t)=0$ (and $h_z(t)=4.0$) with periodic boundary conditions.
 In all cases $L=12$. A reference line corresponding random state entropy is also shown. }
\label {fixedopen}
\end{figure}
As the interaction term in the quantum propagator becomes identity at $\tau=\pi/2$, and indeed any integer multiple of $\pi/2$,
at these values of the kick-period no entanglement is created or destroyed. 

This is reminiscent of the phenomenon of ``quantum resonance'' that 
has been extensively studied for kicked systems such as the kicked-rotor. In that case at specific values of the effective Planck constant that depends on 
the time between kicks the kinetic energy term becomes ineffective, the associated propagator becoming identity, with the consequence that  the energy of the 
kicked rotor increases ballistically in time rather than getting localized \cite{qresonTheo}. This purely quantum phenomenon has also been observed in cold-atom experiments \cite{qresonExp}.  In the present instance also the ineffectiveness of the interaction occurs 
for specific relations between the two time scales, one set by the interaction and the other the time between the pulses of the transverse magnetic field. However the effect is most visible in quantities like entanglement, as the lack of interaction leads to no generation of such quantum correlations.
 
\begin{figure}[!t]
\includegraphics [angle=0,width=1.0\columnwidth] {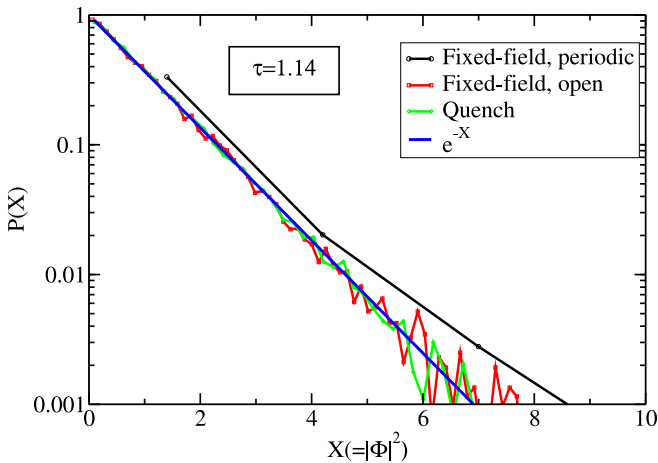}
\caption{ The probability distribution of intensities $X=\vert \br i |\phi_n\kt \vert^2$ for fixed-field kicking cases with periodic and open boundary conditions 
and quench case with periodic boundary conditions are shown for $\tau=1.14$. 
 In all the cases $L=12$. For fixed-field cases $h_x=2.0$ and $h_z=2.0$, and for quench case $h_x(t)$ (and $h_z(t)$) starts from $h_x(t)=4.0$ (and $h_z(t)=0$) and 
finishes at $t_{\rm{max}}$ to $h_x(t)=0$ (and $h_z(t)=4.0$).
For a reference, the exponentially distributed random states case is also shown.} 
\label {intens_dist}
\end{figure}
An interesting case is for large $\tau$ and other than multiples of $\pi/2$. The von Neumann entropy corresponding to these values is equal to 
the entropy of typical or random states. As conjectured by Page \cite{page}, and later proved by others \cite{Sen}, the average entropy of a subsystem of 
dimension $m\leq n$, for pure random states in the $mn$ dimensional space  is given as $\sum_{k=n+1}^{mn}\frac{1}{k}-\frac{m-1}{2n} \approx
\ln m-\frac{m}{2n}$. The approximation is valid for large $m$ and $n$. This shows the somewhat remarkable result that most of the states of pure bipartite 
systems are nearly maximally entangled, the maximum possible being $\ln m$.

In the present case $m=n=2^6$, hence the value turns out to be approximately $5.27$. From Fig. \ref{entropy_rate}
one finds that for $\tau$ not very small and also excepting special values corresponding to $\tau=\pi/2,\pi,3\pi/2,...$, and their immediate vicinity, 
the entropy reaches the average
random state value. In the absence of longitudinal field, the entropy at $h_x=0$ never reaches to the random state value. The longitudinal field and 
the resultant nonintegrability generates the large entanglement.

In  order to get more insight for the random states in a simple spin chain problems, we compare the case with a constant field
kicking case. For this protocol, we apply constant transverse and longitudinal magnetic fields. 
In this case, the unitary operators $U_z$
and $U_x$ take the form
  \begin{eqnarray}
 U_z=\prod_{j=1}^{L_0} \exp\bigl(-i\tau \sigma_j^z\sigma_{j+1}^z-h_z^{\rm fix}\tau \sigma_j^z\bigr),
\end{eqnarray}
and
\begin{eqnarray}
 U_x=\prod_{j=1}^{L_0} \exp\bigl(-i\tau h_x^{\rm fix}\tau \sigma_j^x\bigr).
\end{eqnarray}
The quantum map is now $U=U_x U_z$, and time evolution propagator for time $t=n \tau$ is simply given by $U^n$. The operator is independent 
of the step $k$ and therefore is the stationary Floquet operator. 
The state at time $n \tau$ is $|\psi_n \kt = U^n \otimes^L |\rarrow \kt$. 
 The fixed-field protocol along with quench case is shown in fig. \ref{fixedopen}. 
For the fixed-field situation we consider both periodic ($L_0=L$) as well as open ($L_0=L-1$) boundary conditions. 
For both these boundary conditions, constant fields $h_x=h_z=2.0$ are applied during the process in contrast to the  quench case 
where the magnitude of $h_x$ and $h_z$ vary at each time. It should be noted here that in the quench case both, periodic and open, boundary conditions are qualitatively similar, hence only the
quench case with periodic boundary conditions is shown.
 The entropy for fixed-field periodic and open cases at $t_{\rm max}$ show a dip at the new series of points $\tau=\pi/4, 3\pi/4, 5\pi/4, \cdots$ as well as 
at the already discussed values of $\tau=\pi/2, \pi,3\pi/2, \cdots$.  Interestingly we can see that for a very simple case with 
constant $h_x$ and $h_z$ and open boundary, excepting few special values of $\tau$ 
like $\alpha t_{\rm max}=\pi/2$, the entropy at $t_{\rm max}$ shows the numerical value corresponding to those of random states. However the same is not true for the case 
of periodic chain, where a slightly smaller value is found. Also the 
autonomous case shows peculiar $U$ shaped curves at multiples of $\pi/4$. 

 A detailed analysis of these interesting oscillations is warranted but is postponed however for future study.

The emergence of random states behavior of $|\psi_n\kt$ can be analyzed further by 
calculating the distribution of 
the intensities of defined as $\vert\br i |\psi_n\kt \vert^2$, 
being the $i^{\rm th}$ component of the state $|\psi_n\kt$. 
  In Fig.~(\ref{intens_dist}) is shown the distribution of intensities for $L=12$ spins
with $\tau=1.14$ 
and $t_{\rm{max}} =400$. In order to reach at the maximum time $t_{{\rm max}}$
we need approximately $350$ kicks. During the time evolution we collect all the states 
$\vert\psi_n\rangle$ between kicks $330$ to $340$. The intensity is calculated for the
ensemble of these $40960$ intensity values.
 Typical or random states have real and imaginary parts that are to an excellent approximation independent Gaussian distributed, hence the intensity follows 
the exponential distribution. We see that the quench case (open and periodic) and the fixed field open chain
follows such an exponential distribution, however the same is not true for the fixed field periodic case,
which shows deviations. These origins of these deviations must be a combination of translational symmetry and the autonomous nature of the system.  
\begin{figure}[!t]
\includegraphics [angle=0,width=1.0\columnwidth] {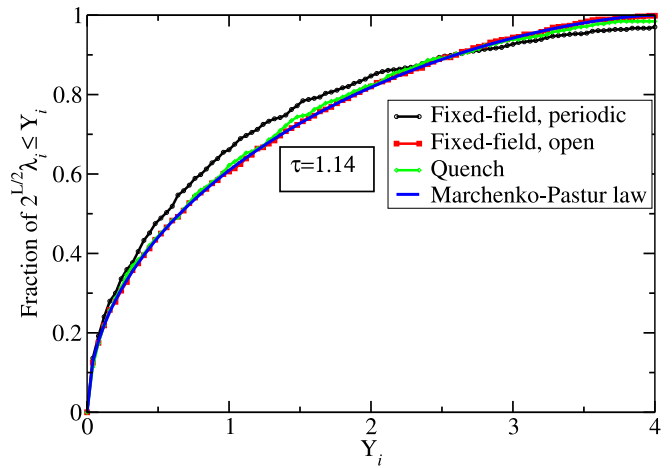}
\caption{Cumulative distribution of eigenvalues of reduced density matrix of $L/2$ spins for $\tau=1.14$ and $L=12$ 
for fixed-field cases with periodic and open boundary conditions and quench case with periodic boundary 
conditions are shown. 
 For fixed-field cases $h_x=2.0$ and $h_z=2.0$, 
and for quench case $h_x(t)$ (and $h_z(t)$) starts from $h_x(t)=4.0$ (and $h_z(t)=0$) and 
finishes at $t_{\rm{max}}$ to $h_x(t)=0$ (and $h_z(t)=4.0$).
For a reference, the Marchenko-Pastur distribution for random states case is also shown.} 
\label {pastur}
\end{figure}

The eigenvalue spectrum of the reduced density matrix is of evident interest and is closely connected to the so-called entanglement spectrum (if $\ld_i$ are the 
eigenvalues, the entanglement spectrum is $-\log \ld_i$). The distribution of eigenvalues of the reduced density matrices of pure random states are known
to follow the Marchenko-Pastur distribution \cite{pastur,karol} for large system dimensionality. If $(\lambda_1,\lambda_2, \cdots, \lambda_{2^{L/2}})$ be the 
eigenvalues of the reduced density matrix for a state of $L$ qubits, the distribution 
for a rescaled variable $Y_i=2^{L/2}\lambda_i$ is given by:
\begin{eqnarray}
 P(Y)=\frac{1}{2 \pi}\sqrt{\frac{4-Y}{Y}},
\end{eqnarray}
   The cumulative density is $P_d(Y)=\int_0^Y  P(x) dx$.
  
  In Fig.~\ref{pastur} this cumulative distribution is plotted for the quench case (periodic), 
the fixed-field open chain case and the fixed-field 
periodic chain. In all cases $L=12$ and $t_{\rm max}=400$. The ensemble of states used is the same as one for the data in Fig.~(\ref{intens_dist}). For each 
state of the ensemble, consisting of pure states of $12$ spins, we calculate all the $2^6$ eigenvalues of the reduced density matrix of $L/2$ contiguous  spins. 
There are $10$ such states calculated at consecutive times, and hence an  
ensemble of $2^6\times 10$ eigenvalues is use for the cumulative density. A reference curve for 
Marchenko-Pastur distribution is also shown. We find that the distribution of eigenvalues in 
quench and fixed-field
open chain cases follow Marchenko-Pastur law while fixed-field periodic chain case shows a deviation.
This is again in agreement with the intensity analysis. Thus it is interesting that quenching encourages the production of states that follow 
properties of random states more closely. 
\section{conclusion}
\label{conclude}
We have studied various entanglement properties
for the non-integrable kicked Ising model with both transverse and longitudinal fields. The 
concurrence, $Q-$measure and von Neumann entropy is calculated numerically for varying longitudinal field. Oscillations feature in the concurrence, 
$Q-$measure and entropy as the transverse field quenches to zero, with a overall decrease in the concurrence and an increase in the other measures 
for small longitudinal fields, signaling the creation of multipartite entanglement. This multipartite entanglement is not produced if the 
longitudinal fields are large enough so that the disordering effects of the transverse field are not felt. 
This paper has also shown the effect of number of kicks (or kicking interval) on entanglement measure such as von Neumann entropy. 
In one protocol we applied sinusoidal transverse and longitudinal time dependent fields and ensured that the rate at which 
transverse field vanishes is same as that at which longitudinal field reaches its maximum value. We find value of 
the kicking period where entanglement exactly vanishes and those where it corresponds to very large 
values that we have identified with those of  random states. The eigenvalues and eigenstates distribution confirm the random state nature of 
generic quenched states in the presence of both transverse and longitudinal fields. For a qualitative comparison, we have also presented 
the case of fixed field kicking using open as well as periodic chains. Further work to elucidate the interesting structure of entanglement in 
these states is underway.    
\section*{Acknowledgments} 
SKM acknowledges the support of DST project ``Quantum chaos and quantum information in condensed matter systems'',
SR/S2/HEP-12/2009 during his stay as a research fellow at IIT Madras where this work was initiated.

\end{document}